\documentclass[fleqn,12pt,epsf,twoside]{article}
\usepackage{espcrc1}

\usepackage{graphicx}
\usepackage[figuresright]{rotating}

\newcommand{\AmS}{{\protect\the\textfont2
  A\kern-.1667em\lower.5ex\hbox{M}\kern-.125emS}}

\title{Generalized Parton Distributions and Constituent Quarks}

\author{ S. Scopetta\address[MCSD] {
Dipartimento di Fisica, Universit\`a degli Studi
di Perugia and INFN, Sezione di Perugia \\
via A. Pascoli
06100 Perugia, Italy} and
V. Vento\address[MCSD] {
Departament de Fisica Te\`orica,
Universitat de Val\`encia
and Institut de F\'{\i}sica Corpuscular,
Consejo Superior de Investigaciones Cient\'{\i}ficas  \\
46100 Burjassot (Val\`encia), Spain } }    
\begin{document}
\def\bra{\langle }
\def\ket{\rangle }

\maketitle

\begin{abstract}
An approach is described to calculate Generalized Parton Distributions
(GPDs) in Constituent Quark Models (CQM).
The GPDs are obtained 
from wave functions to be evaluated
in a given CQM. 
The general relations linking the twist-two GPDs to
the form factors and to the leading
twist quark densities are recovered.
Results for the leading twist, unpolarized
GPD in the Isgur and Karl model are presented.
\end{abstract}

\section{INTRODUCTION}

Generalized Parton Distributions (GPDs) are becoming
one of the main topics of interest in hadronic physics
\cite{jig}. 
GPDs are a natural bridge between exclusive processes,
such as elastic scattering, described in terms of form factors,
and inclusive ones, described in terms of structure functions.
As it happens for the usual Parton Distributions (PDs), the measurement of
GPDs provides us with a unique way to access several 
features of the structure of the nucleon, such as
the quark orbital angular momentum contribution to the proton spin
\cite{ji1,jaffe}.
Therefore,
relevant experimental efforts to measure GPDs will
take place in the next few years 
and it becomes urgent to produce 
predictions for these quantities.
Several calculations have been already performed by using different 
approaches \cite{jig,meln}
and an impressive effort has been devoted to study
their QCD evolution properties \cite{scha2}.

A step towards calculations of GPDs  
in Constituent Quark Models (CQM) can be found in \cite{burk1},
and a consistent approach has been proposed in \cite{ssvv}. 
The CQM has a long story of successful
predictions in low energy studies of the 
structure of the nucleon.
In the high energy sector,
in order to compare model predictions
with data, 
one has to evolve, according to QCD, the leading twist
component of the physical structure functions obtained
at the low momentum scale associated with the model.
Such a procedure, already addressed in \cite{pape}, 
has proven
successful in describing the gross features of 
standard PDs 
by using different CQM
(see, e.g., \cite{trvv}).
Similar expectations motivated the study of GPDs described in
\cite{ssvv}, where
a simple formalism is described to calculate
GPDs from any model.
In this talk, the approach of \cite{ssvv} is reviewed and
applied to the Isgur and Karl (IK) \cite{ik} model.

\section{GENERAL FORMALISM}

Let us think to 
diffractive DIS off a nucleon target,
with initial
and final momenta $P$ and $P'$, respectively. 
GPDs describe the amplitude for finding a quark with momentum fraction
$~~x+\xi/2$ (in the IMF) in the nucleon and replacing it back into
the nucleon with a momentum transfer $\Delta^\mu$.
The GPD $H(x,\xi,\Delta^2)$
is introduced by defining the twist-two part
of the light-cone correlation
function \cite{ji1}
\begin{eqnarray}
\label{eq1}
\int {d \lambda \over 2 \pi} e^{i \lambda x}
\bra P' | \bar \psi (- {\lambda n \over 2})
\gamma^\mu \psi ({\lambda n \over 2}) | P \ket & = & 
\nonumber
\\
= H(x,\xi,\Delta^2) \bar U(P') \gamma^\mu U(P) & + &
E(x,\xi,\Delta^2) \bar U(P') 
{i \sigma^{\mu \nu} \Delta_\nu \over 2M} U(P) + ...
\end{eqnarray}
\\
where 
ellipses denote higher-twist contributions,
$\psi$ is a quark field and M is the nucleon mass.
The $\xi$ variable, the so called ``skewedness'', is defined
by the relation $\xi = - n \cdot \Delta$,
where $n=(1,0,0,-1)/
(2 \Lambda)$ and $\Lambda$ depends on the reference
frame.
The $\xi$ variable is bounded by 0 and 
$\sqrt{-\Delta^2}/\sqrt{M^2-\Delta^2/4}$. 
Besides, one has $t=\Delta^2=
\Delta_0^2 - \vec{\Delta}^2$. 
When the longitudinal momentum fraction of
the quark is less than $-\xi/2$, GPDs describe antiquarks;
when it is larger than $\xi/2$, they describe quarks;
when it is between $-\xi/2$ and $\xi/2$, they describe 
$q \bar q$ pairs.
There are two natural limits for $H(x,\xi,\Delta^2)$: 
i) when $P^\prime=P$, i.e., $\Delta^2=\xi=0$, the so called
``forward'' limit, one recovers
the usual PDs
\begin{eqnarray}
H(x,0,0)=q(x)
\end{eqnarray}
ii)
the integration over $x$ yields the Dirac 
Form Factor (FF)
\begin{eqnarray}
\int dx H(x,\xi,\Delta^2) = F_1(\Delta^2)~. 
\end{eqnarray}
Any model estimate of the GPDs has to respect the above two 
crucial constraints.

In Ref. \cite{ssvv},
the Impulse Approximation (IA) expression
for the GPD 
$H(x,\xi,\Delta^2)$, suitable to perform CQM
calculations, has been obtained.
In particular it has been found that,
substituting the quark fields in the left-hand-side of Eq.(1), 
taking into account the quarks degrees of freedom only,
using IA, considering
a process with
${\vec \Delta}^2 \ll M^2$
in the Nucleon rest frame,
using  a symmetric wave function
(as the one given in a NR quark model
once color has been taken into account), one obtains
\begin{eqnarray}
H(x,\xi,\Delta^2)  = \int d \vec k\,\,\,
\delta \left (x + { \xi \over 2} - {k^+ \over M}  \right )\,
\tilde n(\vec k , \vec k + \vec \Delta)~, 
\end{eqnarray}
where $\tilde n(\vec k , \vec k + \vec \Delta)$ is
the one-body non-diagonal momentum distribution:
\begin{eqnarray}
\tilde n(\vec k , \vec k + \vec \Delta)
& = & 
3 \int \psi^*(\vec k_1, \vec k_2, \vec k + \Delta)
\psi(\vec k_1, \vec k_2, \vec k) d \vec k_1 d \vec k_2 =
\nonumber
\\
& = & \int e^{i ((\vec k + \vec \Delta) \vec r
-\vec k \vec r' )} \rho(\vec r, \vec r') d \vec r
d \vec r'~,
\end{eqnarray}
defined through the one-body non diagonal charge density
\begin{eqnarray}
\rho(\vec r, \vec r') 
=  
\int \psi^*(\vec r_1, \vec r_2, \vec r')
\psi(\vec r_1, \vec r_2, \vec r) d \vec r_1 d \vec r_2 ~.
\end{eqnarray}

\begin{figure}[htb]
\begin{minipage}[t] {77 mm}
\vspace{7.7cm}
\includegraphics{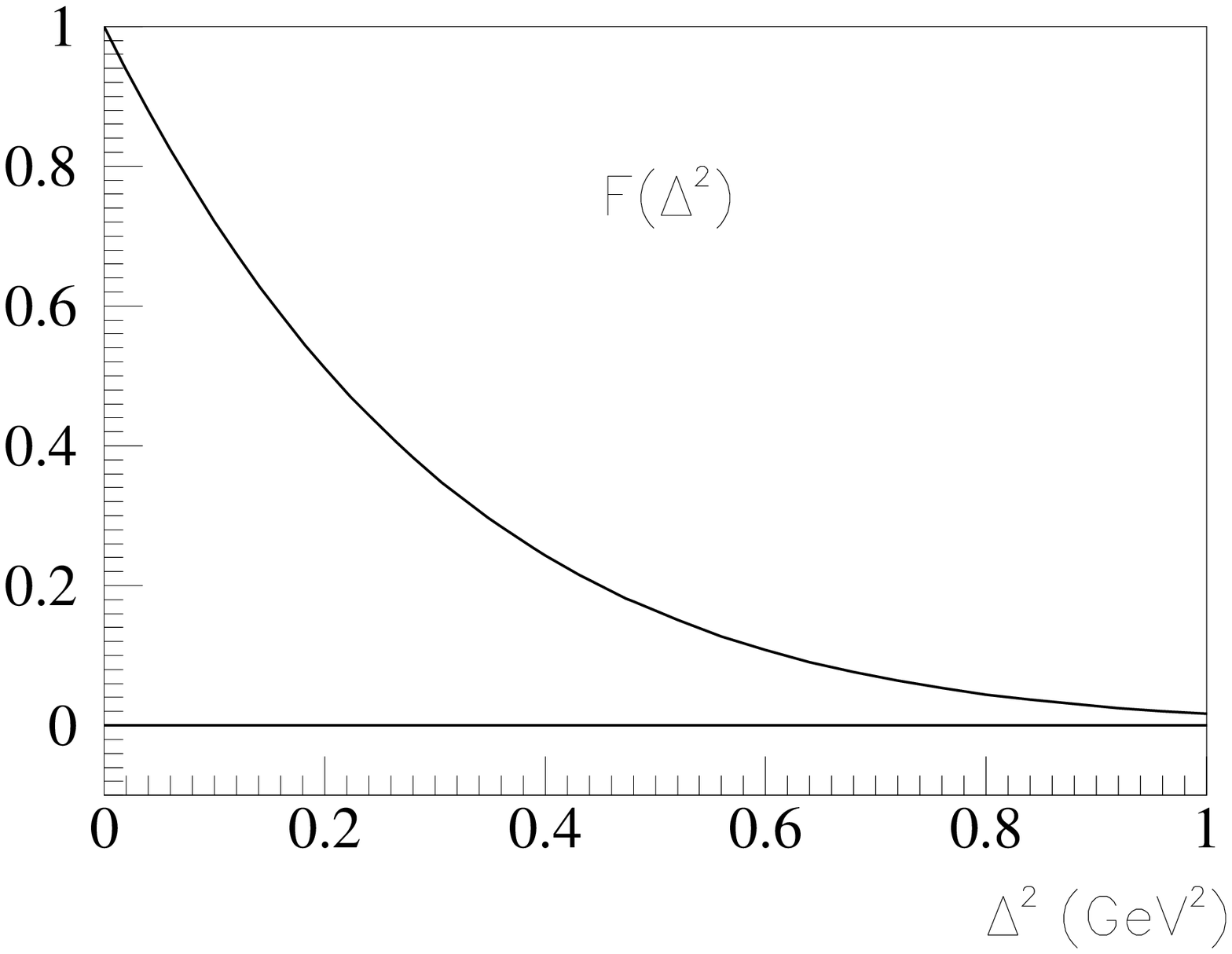}
\vskip -2.5cm
\caption{The charge FF in the IK model.}
\end{minipage}
\hspace{\fill}
\begin{minipage}[t] {77 mm}
\vspace{7.7cm}
\includegraphics{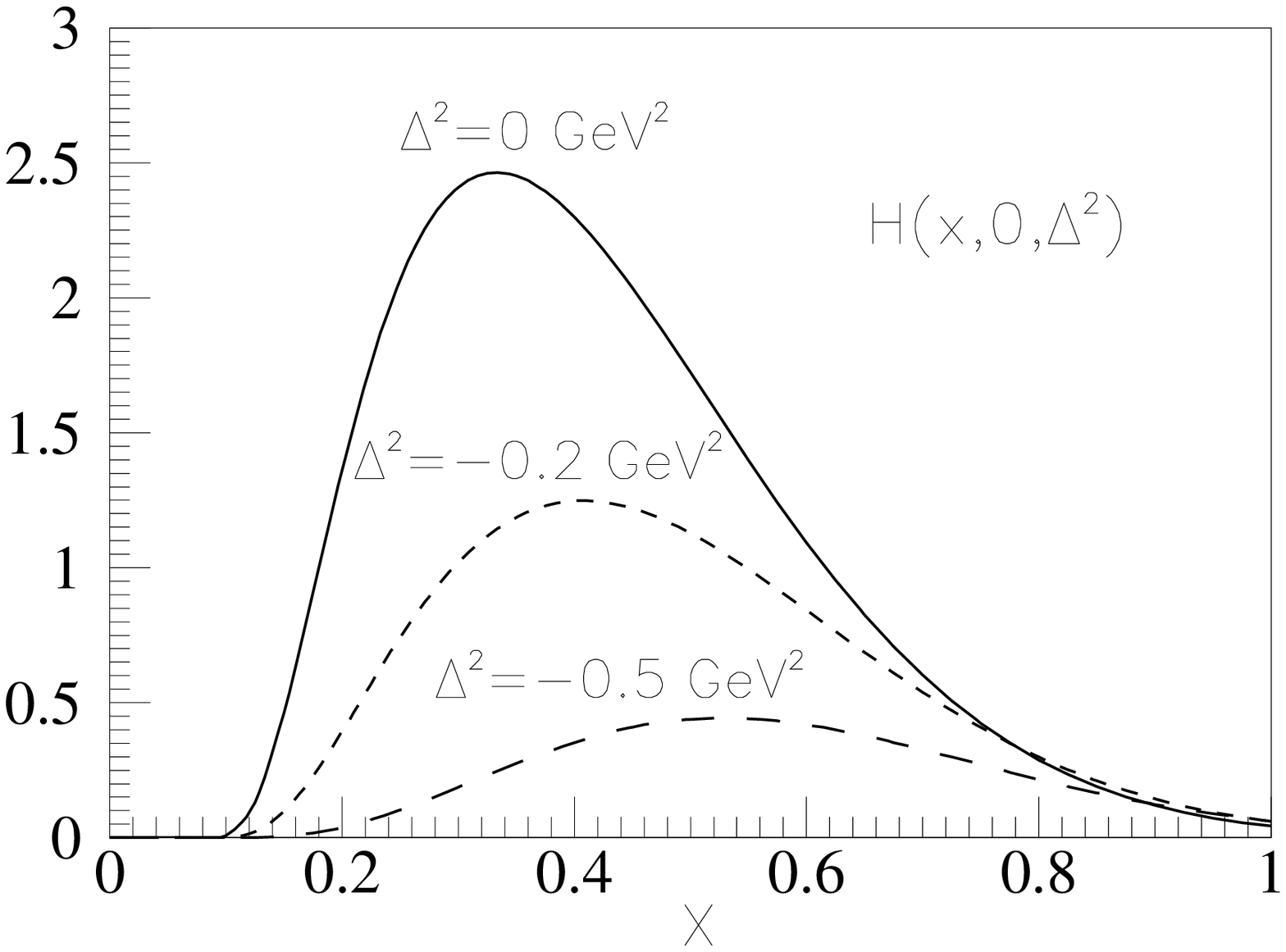}
\vskip -2.5cm
\caption{The GPD $H(x,\xi,\Delta^2)$ at $\xi=0$ and three values of
$\Delta^2$.}
\end{minipage}
\end{figure}

Eq. (4)
allows the calculation of the GPD
$H(x,\xi,\Delta^2)$ in any CQM, and
it naturally verifies the two crucial constraints, Eqs. (2) and (3)
\cite{ssvv}. 
With respect to Eq. (4),
a few caveats are necessary.
Due to the use of CQM wave functions, 
only quarks GPDs can be evaluated, i.e.,
only the region $x \geq \xi/2$ 
can be explored.
The approach has to be improved 
in order to study the sea region
($ - \xi / 2 \leq x \leq \xi/2$).
Besides,
in the argument of the $\delta$ function
in Eq. (4), due to the used approximations, the $x$ variable
is not defined in its natural support, i.e. it can be
larger than 1.
Although the support violation is
small in most models, this problem has to be considered
a serious drawback of all CQM calculations of parton distributions,
in particular if pQCD evolution of the model prediction
is performed.
We stress that our definition of the GPD $H(x,\xi,\Delta^2)$
can be easily generalized
to other GPDs, which can be obtained
in any CQM.

\section{RESULTS IN THE ISGUR AND KARL MODEL}

We consider
the Isgur-Karl model \cite{ik}, with a proton wave function 
given by a harmonic oscillator
potential 
including contributions up to the $2 \hbar \omega$ shell. 
In this case the proton state is given by the
following admixture of states

\begin{equation}
|N \rangle = 
a_{\cal S} | ^2 S_{1/2} \rangle_S +
a_{\cal S'} | ^2 S'_{1/2} \rangle_{S} +
a_{\cal M} | ^2 S_{1/2} \rangle_M +
a_{\cal D} | ^4 D_{1/2} \rangle_M~,
\label{ikwf}
\end{equation}
where we have used the spectroscopic notation $|^{2S+1}X_J \rangle_t$, 
with $t=A,M,S$ being the symmetry type.
The coefficients were determined by spectroscopic properties to be: 
$a_{\cal S} = 0.931$, 
$a_{\cal S'} = -0.274$,
$a_{\cal M} = -0.233$, $a_{\cal D} = -0.067$.

\begin{figure}[htb]
\begin{minipage}[t] {77 mm}
\vspace{7.7cm}
\includegraphics{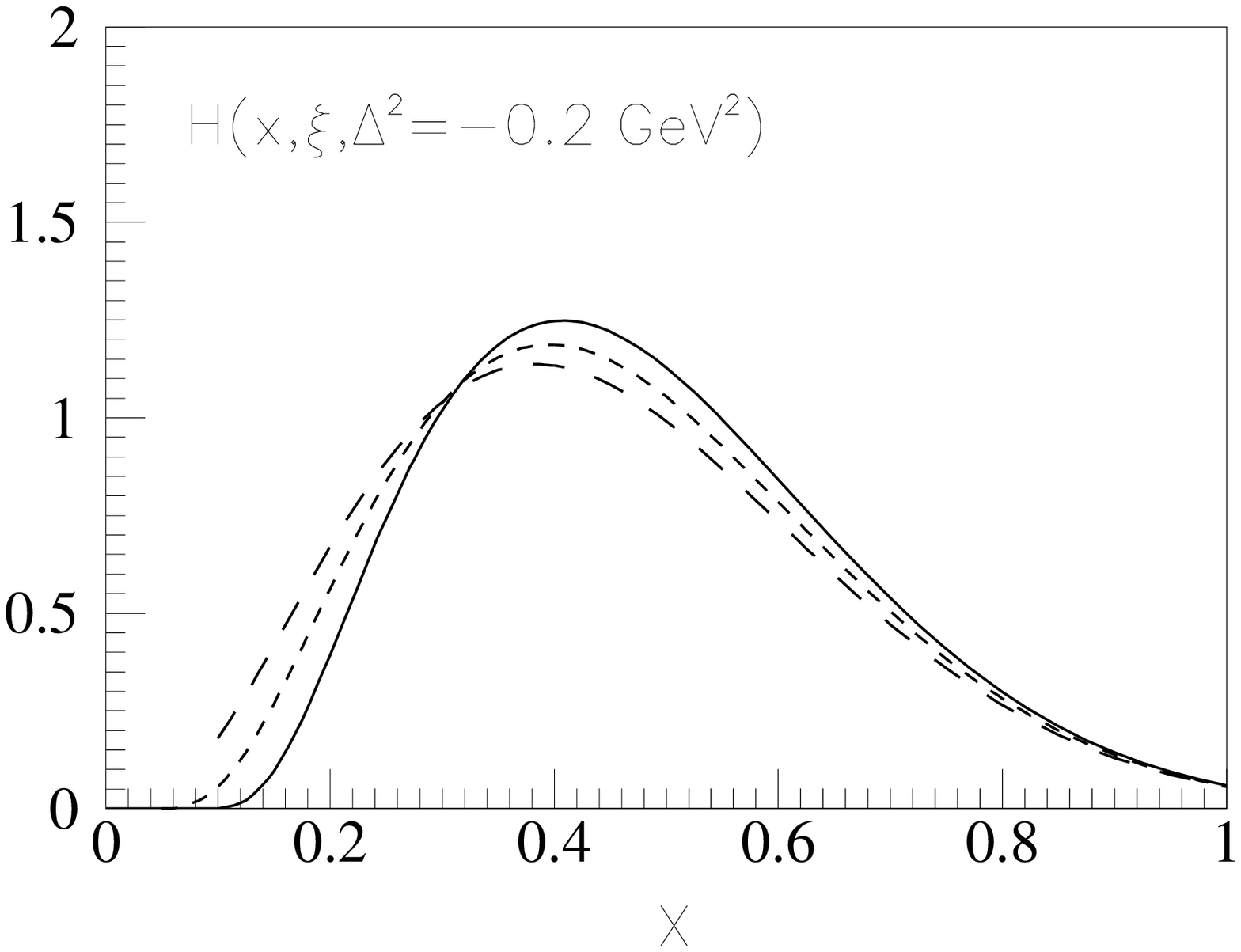}
\vskip -2.5cm
\caption{The GPD $H(x,\xi,\Delta^2)$ at $\Delta^2=-0.2$ GeV$^2$
and $ \xi =0 $ (full line),  $ \xi =0.1 $ (dotted line),
$ \xi =0.2 $ (dashed line). Notice that $H(x,\xi,\Delta^2)$
is shown for $x \geq \xi/2$. }
\end{minipage}
\hspace{\fill}
\begin{minipage}[t] {77 mm}
\vspace{7.7cm}
\includegraphics{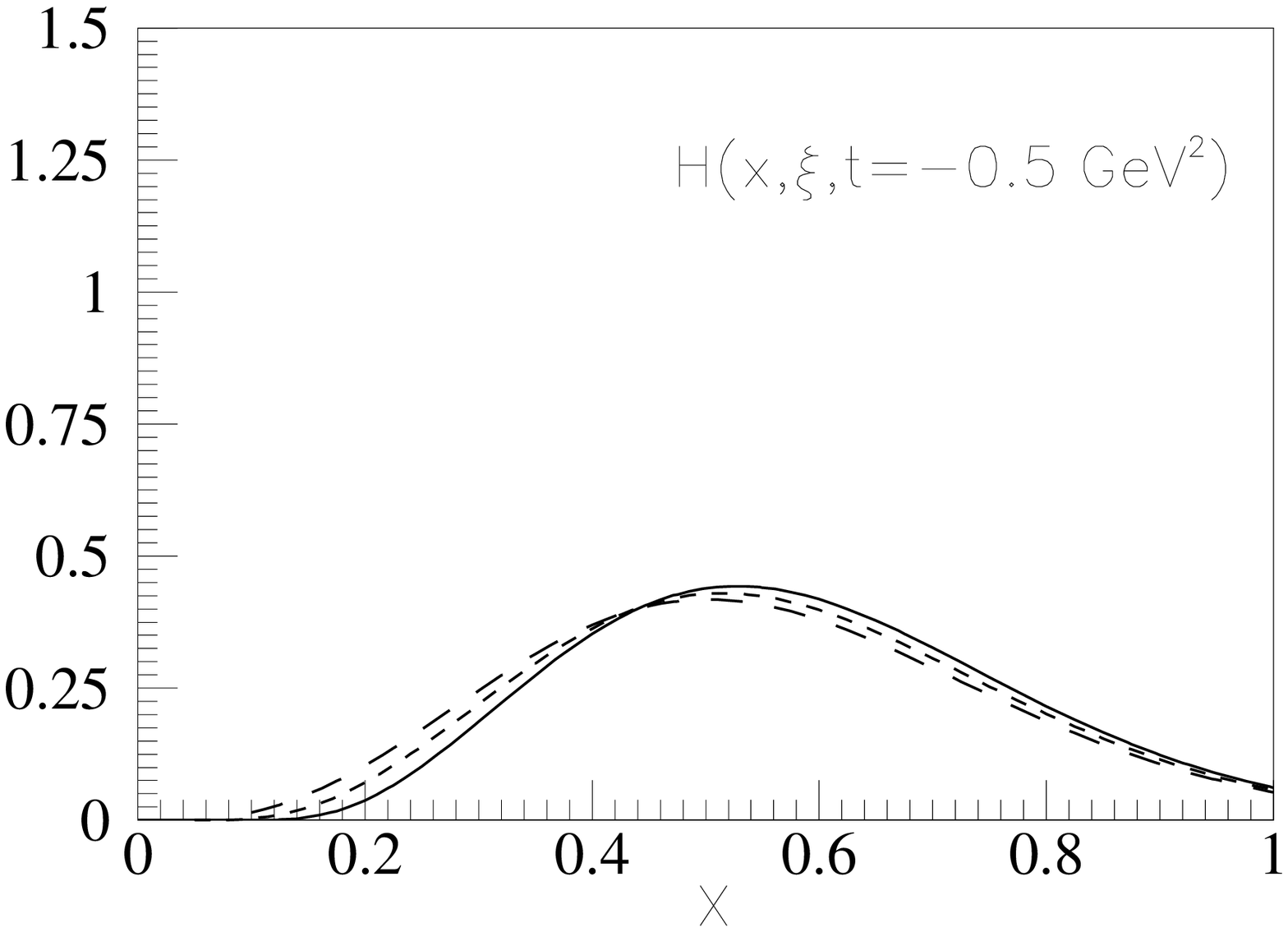}
\vskip -2.5cm
\caption{ The GPD $H(x,\xi,\Delta^2)$ at $\Delta^2=-0.5$ GeV$^2$
and $ \xi =0 $ (full line),  $ \xi =0.1 $ (dotted line),
$ \xi =0.2 $ (dashed line). Notice that $H(x,\xi,\Delta^2)$
is shown for $x \geq \xi/2$. }
\end{minipage}
\end{figure}

The obtained behavior of the FF is shown in Fig. 1.
As it is well known, such a FF
underestimates
the data at high $t$. 
Results evaluated using the IK model, Eq. (7),
in the general formula, Eq. (4), are shown in Figs. 2 to 4.
In Fig. 2, we show the $t$ dependence of our results.
The full line corresponds to the usual PD.
One immediately realizes that a strong $t$ dependence is found,
in comparison with other estimates, for example, with the one
in \cite{meln}. This has to do with the
too a strong $t$ dependence 
of the FF in the IK model.
In Figs. 3 and 4 we have the full $t$ and $\xi$ dependences.
These are similar to the ones obtained in \cite{meln},
although the $\xi$ dependence is stronger.

Our results for $H(x,\xi,\Delta^2)$ correspond to
the low momentum scale associated with the model. 
In order to compare them with the data which are going to be taken
in future experiments, one has to evolve them to the
experimental high-momentum scale.
 
The proposed approach 
can have many interesting developments,
such as the use of
more realistic models, the inclusion of
QCD evolution
from the scale of the model to the experimental one,
the addition of corrections due to relativistic
dynamics, or the ones due to a 
possible finite size and complex structure of the
constituent quarks, as proposed by several authors \cite{acmp}.

\end{document}